\documentclass[12pt,manuscript]{aastex}
\newcommand{\be}{\begin{equation}}
\newcommand{\ee}{\end{equation}}
\newcommand{\bea}{\begin{eqnarray}}
\newcommand{\eea}{\end{eqnarray}}
\shorttitle{Very Hot Planets}
\shortauthors{C. Beaug\'e and D. Nesvorn\'y} 
\begin{document}
\title{Emerging Trends in a Period-Radius Distribution of Close-in Planets}
\author{C. Beaug\'e$^{1,2}$ and D. Nesvorn\'y$^2$}
\affil{(1) Instituto de Astronom\'{\i}a Te\'orica y Experimental (IATE), Observatorio Astron\'omico, \\
Universidad Nacional de C\'ordoba, Laprida 854, X5000BGR C\'ordoba, Argentina}
\affil{(2) Department of Space Studies, Southwest Research Institute, \\ 
1050 Walnut St., Suite 300, Boulder, CO 80302, USA}

\begin{abstract}
We analyze the distribution of extrasolar planets (both confirmed and {\it Kepler} candidates) according to their orbital periods $P$ and planetary radii $R$. Among confirmed planets, we find compelling evidence for a paucity of bodies with $3 < R < 10 R_\oplus$, where $R_\oplus$ in the Earth's radius, and $P < 2$-$3$ days. We have christened this region a {\it sub-Jovian Pampas}. The same trend is detected in multiplanet {\it Kepler} candidates. Although approximately 16 {\it Kepler} single-planet candidates inhabit this Pampas, at least 7 are probable false positives (FP). This last number could be significantly higher if the ratio of FP is higher than $10 \%$, as suggested by recent studies. 

In a second part of the paper we analyze the distribution of planets in the $(P,R)$ plane according to stellar metallicities. We find two interesting trends: (i) a lack of small planets ($R < 4 R_\oplus$) with orbital periods $P < 5$ days in metal-poor stars, and (ii) a paucity of sub-Jovian planets ($4 R_\oplus < R < 8 R_\oplus$) with $P < 100$ days, also around metal-poor stars. Although all these trends are preliminary, they appear statistically significant and deserve further scrutiny. If confirmed, they could represent important constraints on theories of planetary formation and dynamical evolution. 

\end{abstract}


\section{Introduction}

Close-in (or Hot) planets, usually defined as those having semimajor axes $a < 0.1$ AU (or orbital periods $P < 10$ days), are the easiest to detect, both with radial velocity (RV) surveys and transits. Almost half of the confirmed planets currently known correspond to this population, although this proportion is certainly affected by observational bias. It is believed that close-in planets cannot have been formed in situ (e.g. Lin et al. 1996), and thus constitute an interesting evidence for orbital migration and dynamical evolution of extrasolar planetary systems. 

While most of the exoplanets detected by Doppler techniques correspond to giant planets (typically, masses $m \ge 0.3 m_{\rm Jup}$), the recent discoveries from {\it Kepler} have been dominated by much smaller planets, usually in the Super-Earth and Neptune mass range. Although this may point to the fact that smaller planetary bodies are more numerous (e.g. Mayor et al. 2009, Howard et al. 2010, Howard et al. 2012), the exact statistics also depends on metallicities of the host stars (e.g. Fischer \& Valenti 2005, Santos et al. 2011).  

The distribution of planets in a planetary radius ($R$) vs. orbital period ($P$) plane provides important information about planetary formation and migration in different planet-size regimes (e.g. Ben\'{\i}tez-Llambay et al. 2011, Latham et al. 2011, Youdin 2011, Hasegawa \& Pudritz 2012). Also, planetary occurrence in the $(P,R)$ plane for different stellar metallicities and effective temperatures $T_{\rm eff}$ may lead to insights on how these parameters affect both planetary formation and orbital migration. For example, there is indication that sub-Jovian planets may be found in a wider range of metallicities than giant planets (Buchhave et al. 2012), and that giant planet occurrence increases with  $T_{\rm eff}$ and stellar mass (Johnson et al. 2010, Howard et al. 2012).

In this paper we perform a detailed analysis of planets in the $(P,R)$ plane, including public data from both confirmed planets and Kepler planetary candidates. We restrict our analysis to planets with orbital periods $P < 50$ days and host stars with masses $m_* > 0.5 m_\oplus$ Our goal is to search for possible (statistically significant) trends in the $(P,R)$ plane and discuss possible explanations for these trends. In Section 2 we analyze the distribution of close-in planets in the $(P,R)$ plane and point out the possible existence of a sub-Jovian desert for orbital periods lower than $\sim 2$-$3$ days. In Section 3 we discuss several new trends in the $(P,R)$ distribution of planets according to the stellar metallicity. For planets without detected transits we extend our analysis to the plane of orbital period vs. minimum planetary mass (i.e. $(P,m)$ diagram). Discussions and possible dynamical interpretations of the detected trends close the paper in Section 5.

\section{The Distribution of Sub-Jovian Planets}

\subsection{A Sub-Jovian Desert?}

\begin{figure}[t]
\centering
\includegraphics[width=0.9\textwidth,clip=true]{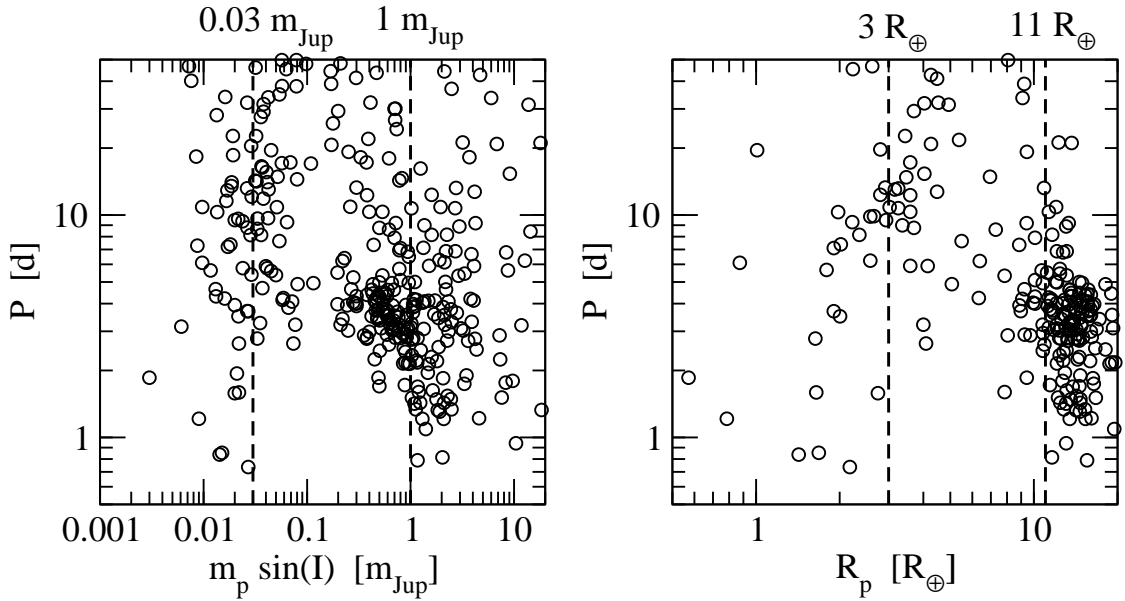}
\caption{Distribution of orbital periods with planet mass (left) and radius (right) for a total of 287 confirmed planets with orbital periods $P < 50$ days.}
\label{fig1}
\end{figure}

Figure \ref{fig1} shows the distribution of orbital periods of all confirmed planets (as of July 2012) with $P < 50$ days, totaling 287. The left plot shows $P$ as a function of the mass, while the one on the right shows the corresponding distribution in terms of the planetary radius $R$. These two data sets are not identical because some planets have no detected transits, and thus no information is known of their radii. 

We can separate the planets, according to their mass, roughly into three groups: Jovian planets ($m \ge 1 m_{\rm Jup}$), Neptunes or Sub-Jovian planets ($0.03 m_{\rm Jup} \le  m < 1 m_{\rm Jup}$), and Super-Earths ($m < 0.03 m_{\rm Jup}$). This division is arbitrary, but it can be useful to highlight different formation mechanisms of different populations.  In terms of the physical radii, these groups can also roughly be defined by the relations: $R \ge 11 R_{\oplus}$ for Jovian planets, $3 R_{\oplus} \le R < 11 R_{\oplus}$ for Neptunes and Sub-Jovian planets, and $R < 3 R_{\oplus}$ for Super-Earths. However, the observed diversity in planetary densities implies that there is no unique relationship between radius and mass, so the above relationship is only illustrative. Also keep in mind that in most cases planetary masses estimated from Doppler surveys are only minimum bounds to actual values, because of the (generally) unknown inclination of the orbital plane with respect to the observer's line of sight.

Even with these reservations in mind, Figure \ref{fig1} shows an apparent absence of Sub-Jovian planets with orbital periods smaller than $P \sim 3$ days. Possibly the first reference to a possible {\it sub-Jovian desert} was made by Szabo \& Kiss (2011), although most of the Super-Earths then detected belonged to small mass stars. Ben\'{\i}tez-Llambay et al. (2011) corrected that distribution by normalizing the semimajor axis by stellar mass and radius; with the exception of CoRoT-7b, the other exoplanet population seemed to have a distribution more in accordance to a {\it step} function. Specifically, the orbital periods of close-in planets with $m > m_{\rm Jup}$ appeared to be restricted to periods $P > 1$ day, while smaller masses seemed to be detected only down to  $P \sim 3$ days. This trend seems compatible with the existence of an inner cavity in the protoplanetary disk acting as a planetary trap for type I migration, plus a long-term evolution due to tidal effects after the gas disk dispersal (Ben\'{
\i}tez-Llambay et al. 2011).

Over the past year, however, as the population of small-mass increased dramatically (especially due to {\it Kepler}), a significant population of small Super-Earth planets has been detected around Solar-type stars with lower orbital periods, also down to $P \sim 1$ day or even lower. Nevertheless, the absence of very hot sub-Jovian planets is still maintained and today, with 287 confirmed planets, the existence of this unpopulated region appears very prominent, especially in the $(P,m)$ plane (Figure \ref{fig1}a).

Although observational bias cannot be ruled out, it seems unlikely. While planetary detection via Doppler techniques favor large masses, several planets have been found in the sub-Jovian mass range with longer orbital periods. Moreover, in principle {\it Kepler} should have little problem in detecting a planet within this proposed sub-Jovian desert. An estimate (Koch 2004) shows that planets in this region should have an SNR value between 400 and 1600 (assuming Solar-type star, observational time of 1 year and an impact parameter $b=0.5$), much higher than most of the confirmed {\it Kepler} planets. 

Population synthesis models (e.g. Ida \& Lin 2004, Mordasini et al. 2009, 2012) predict a paucity of Neptune-size planets relatively close to the star as a result of the interplay between planetary formation timescales and different migration regimes (Type I and Type II). However, these predictions have not been validated (e.g. Howard et al. 2012). The desert proposed by Szabo \& Kiss is too sharply defined in the mass range, includes masses almost up to $1 m_{\rm Jup}$, and is restricted to very small orbital periods. 

As of July 2012, there are more than 2300 {\it Kepler} candidate planets that have not been confirmed nor validated. We tested whether the distribution of these candidates in the $(P,R)$ plane also shows the sub-Jovian desert or introduces a more smoothed distribution. However, we must keep in mind that among {\it Kepler} candidates there are bound to be a number of false positives (FP), whose number is still a matter of debate. Lissauer et al. (2012) argued that although FP may be more common among KOI with only a single transit signal, among targets displaying multiple-planet transits the fraction of real systems could be as large as $95 \%$. 

\begin{figure}[t]
\centering
\includegraphics[width=0.9\textwidth,clip=true]{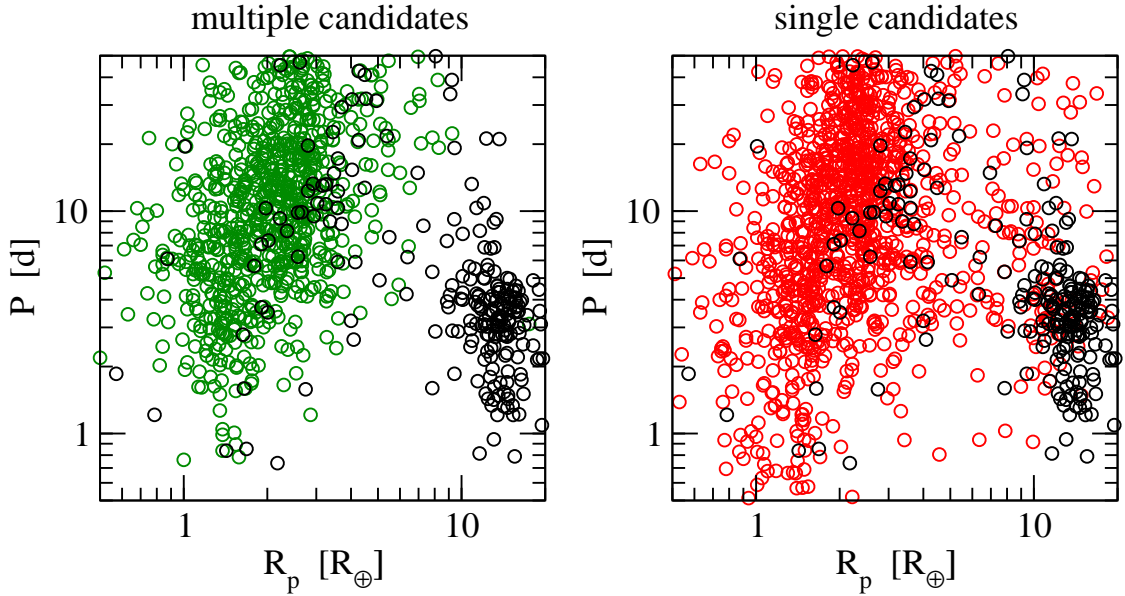}
\caption{Distribution of orbital periods with planet radius, for all confirmed planets (black, $N=232$) and {\it Kepler} candidates (green \& red), separated according to multiple {\it Kepler} systems (left, $N=784$) and single {\it Kepler} candidates (right, $N=1114$).}
\label{fig2}
\end{figure}

The left plot in Figure \ref{fig2} shows the distribution of these candidate systems (green circles), once again in the $(P,R)$ plane. For comparison, the confirmed planets, in great majority coming from observations other than {\it Kepler}, are shown in black. {\it Kepler} multiple candidates tend to be smaller and contain almost no Hot Jupiters. On the other hand, a significant number of Super-Earth candidates are seen, some of them in systems with up to six planets. However, the sub-Jovian desert is still apparent in this data. For $P < 3$ days, there is a large number of Hot Jupiters from RV surveys and a large number of Super-Earths (and Neptunes) from {\it Kepler}, but almost no planets with radii between $3$ and $11 R_\oplus$. 

In order to analyze whether this lack of planets is statistically significant, we performed a very simple Monte Carlo test. We counted the number of detected bodies (including both confirmed planets and multiple systems candidates) with $R \in [3,10] R_\oplus$ and $P \in [0.5,P_{\rm max}]$ days, $P_{\rm max}$ being an upper limit which was varied in successive trials. For each value of $P_{\rm max}$ we then generated a series of $10^6$ fictitious populations with the same number of data points within the same intervals of $P$ and $R$, and counted what percentage of them included no values within the proposed desert. We varied $P_{\rm max}$ between $10$ and $50$ days, and considered uniform distributions in $(P,R)$ as well as in $(\log{P},\log{R})$. Depending on the value of $P_{\rm max}$ and the chosen distribution function, we found that the probability of reproducing the desert was at most $2 \%$, although in most cases much smaller than $1 \%$. Although this test is far from conclusive, its results are 
suggestive. 

\begin{figure}[t]
\centering
\includegraphics[width=0.65\textwidth,clip=true]{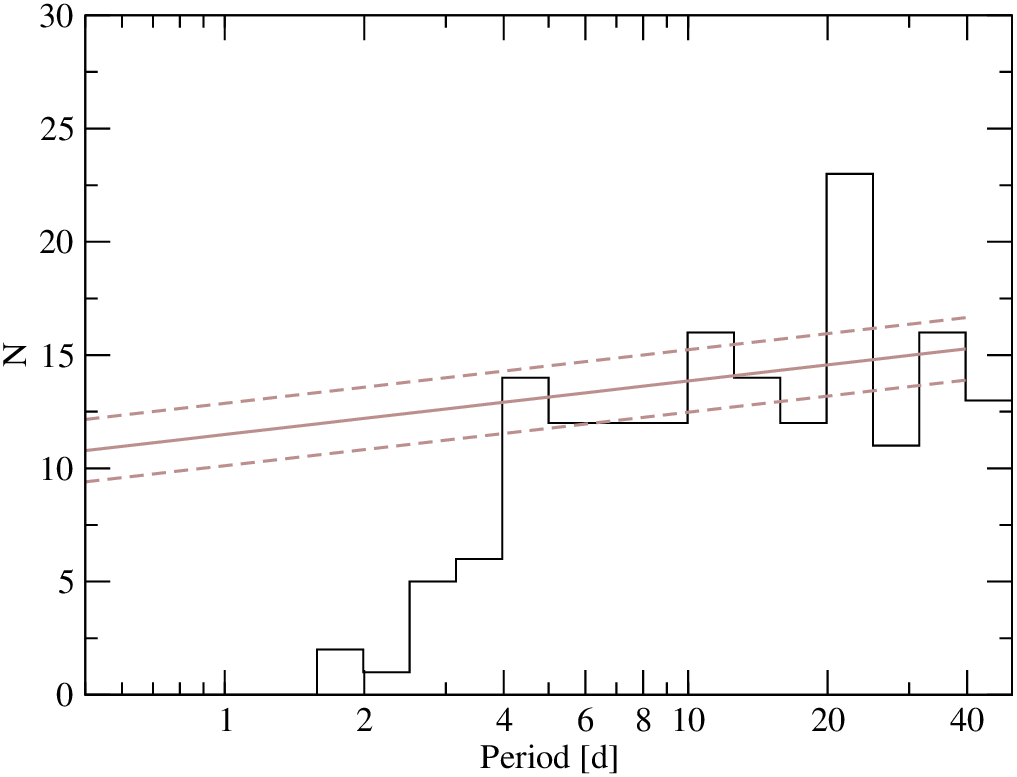}
\caption{Distribution of medium-size planets ($R \in [3,10] R_\oplus$) s function of their orbital period, including both confirmed and {\it Kepler} multi-planet candidates. The gray continuous line shows a linear fit in $\log(P)$ obtained considering only those planets with $P>4$ days. Dashed lines show the $1 \sigma$ values.}
\label{fig3}
\end{figure}

We then proceeded to do a different, and slightly more elaborate statistical test. First, we searched for a region around the suspected desert with a population of detected planets as uniform and homogeneous as possible. Although our first choice was to analyze the distribution in planetary radius around $P \sim 3$ days, we noted that both the stellar populations and detection techniques that dominate each side of the desert are different ({\it Kepler} for smaller planets, and RV for Jovian masses), and therefore it would not be correct to construct a single fitted distribution function for both sub-populations. In consequence, once again we chose the distribution of observed planets according to orbital period, delimited in values of $R$ within $R \in [3,10] R_\oplus$. 

We then binned the observed population of planets in the interval $P \in [0.5,50]$ days and 
fitted the data using a polynomial distribution. We found that the resulting functional fit was practically linear in $\log{P}$, with coefficients that were very robust with respect to the bin size. Together with the coefficients of the fit we also estimated their uncertainties $\sigma_i$. Finally, we projected this fitted distribution function to the region of the proposed desert (i.e. $P < 3$ days) and compared it with the observed distribution. We found that the difference between them is of excess of $7 \sigma$, where $\sigma$ is an estimation of the variance of the fit in this region. An example is shown in Figure \ref{fig3}. Again, the observed lack of planets close to the star does not appear consistent with the distribution found for larger orbital periods.

\subsection{Problematic Cases and Possible FPs}

Although the distribution of {\it Kepler} multi-planet candidates preserves the alleged desert, once the single planet candidates are introduced, the distribution for $P < 3$ days becomes more fuzzy (red circles in Figure \ref{fig2}b). In particular, the sub-Jovian region with $P < 3$ days now appears populated with around 16 planetary candidates. The question therefore arises: does this mean that the sub-Jovian desert is not completely void of planets, or are these ``problematic'' candidates false positives?

Among single candidates, the percentage of FP is expected to be higher than for multiple-candidate systems. Morton \& Johnson (2011) estimated FPs of the order of $10 \%$, while Borucki et al. (2011) mentioned values as large as $20 \%$ for rank 2 KOI, and even $40 \%$ for ranks 3 and 4. More recently, Santerne et al. (2012) performed radial velocity observations on a sample of 46 {\it Kepler} candidates with orbital periods below 25 days, and concluded that as much as $35 \%$ of the single planet candidates could be FPs. Colon et al. (2012) have proposed an even larger FP fraction, close to $50 \%$, especially for small orbital periods. 

Recently, Bonomo et al. (2012) compared the number of planets with $2 R_\oplus \le R \le 4 R_\oplus$ detected by {\it CoRoT} with the number of planets+candidates proposed from {\it Kepler} data. They pointed out that, according to the planetary occurrence ratio proposed by Howard (2012), {\it CoRoT} should have detected a much larger number of Hot Neptunes than actually found. Although the discrepancy could be due to the different stellar populations observed by both missions, it could also be indicative of an underestimation of the FP probability assumed by Howard (2012).

Given these results, it is perhaps possible that the $\sim 16$ candidates within the proposed desert are in fact FPs. We discuss this possibility in more detail below.

Two of our problematic cases (KOI 64.01 and KOI 102.01) were mentioned by Borucki et al. (2011) as possible FPs. Ofir \& Dreizler (2012) presented an independent planet search in the {\it Kepler} data base, using a modified version of the SARS pipeline (Ofir et al. 2010) developed for {\it CoRoT}. The treatment of the data was slightly different than the software used by the {\it Kepler} team. The authors rejected 11 KOIs as eclipsing binaries (EBs) based on close inspection of the light curves. Among these appears another of our problematic cases (KOI 1459.01), which is identified as an EB. Last of all, Col\'on et al. (2012) perform multi-color transit photometry on 4 {\it Kepler} candidates, and find that 2 are indeed false positives. KOI 1187.01, another of our problematic cases, is among them. 

\begin{figure}[t]
\centering
\includegraphics[width=0.95\textwidth,clip=true]{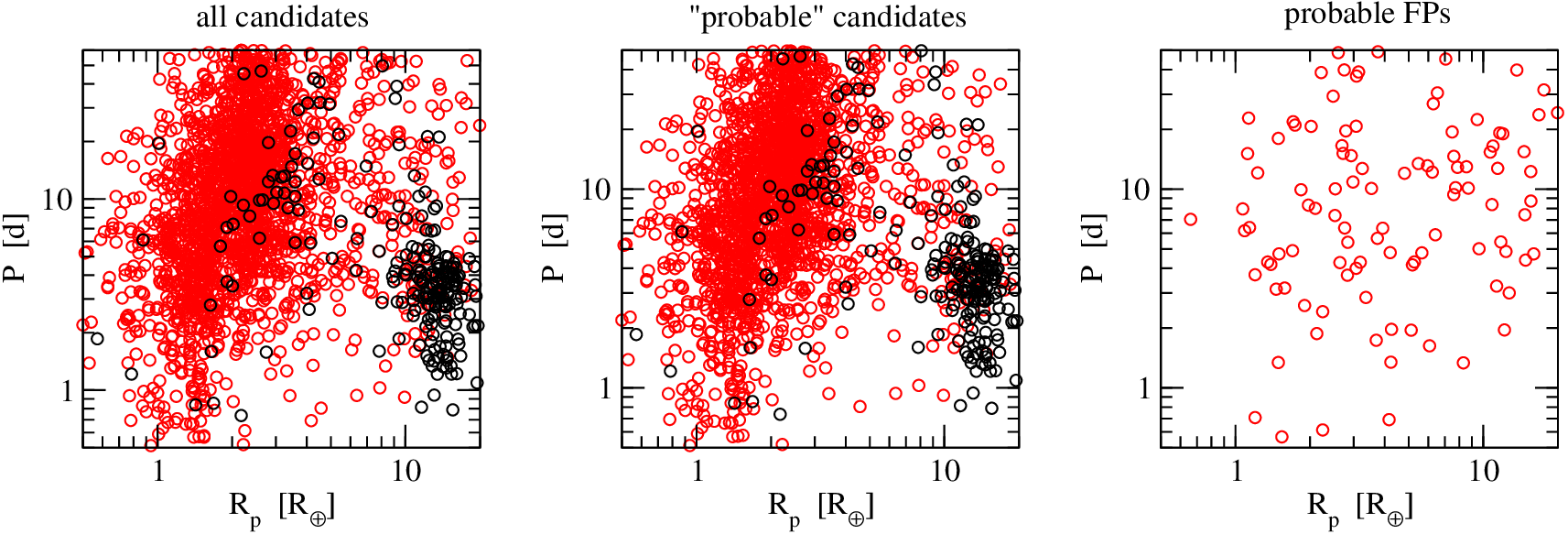}
\caption{Distribution in $(P,R)$ plane of all confirmed planets (black) and {\it Kepler} candidates (red). Plots show all planetary candidates (left), ``probable'' planets (center), and probable False Positives (right). See text for details.}
\label{fig4}
\end{figure}

\begin{figure}[t]
\centering
\includegraphics[width=0.95\textwidth,clip=true]{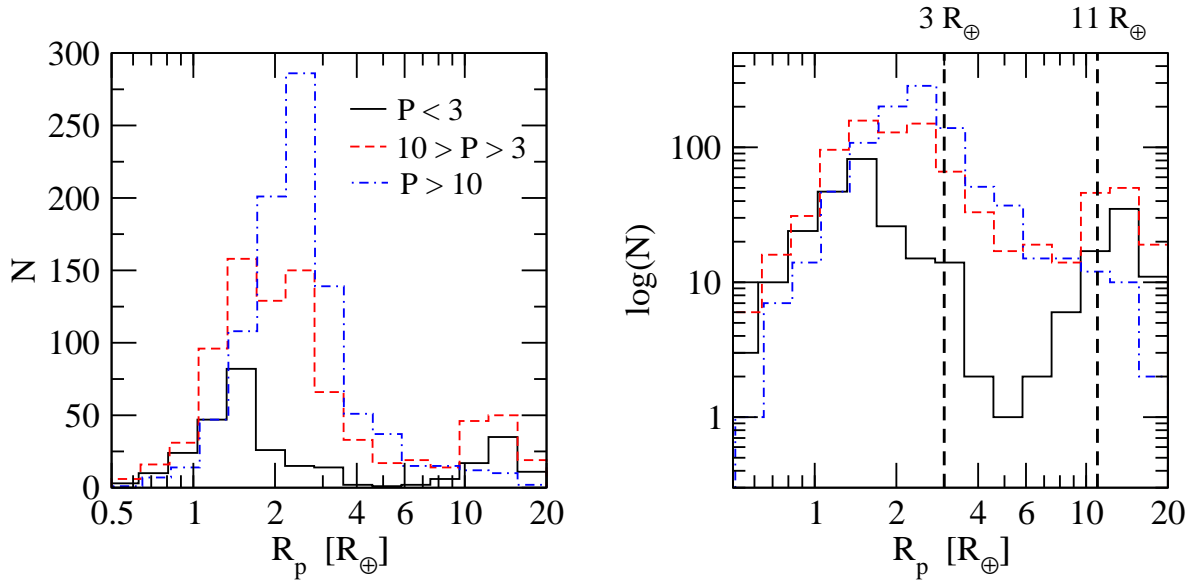}
\caption{Distribution of planets according to radii for three intervals of orbital period ($P < 3$ days in black, $3 \le 3 P < 10$ days in red and $P \ge 10$ days in blue). Plot on the right shows the same data as the one on the left, but with the number of bodies in log scale.}
\label{fig5}
\end{figure}

From these sources we can construct a data set of questionable candidates, subtract them from the planetary candidate list, and thus define a more ``probable'' list of planetary candidates. Figure \ref{fig4} shows the change in the $(P,R)$ distribution of {\it Kepler} candidates when these questionable cases are eliminated. The overall shape of the distribution is maintained, but now the number of planets in the proposed desert has decreased significantly, from 16 to 9. Their KOI numbers are: 356.01, 439.01, 506.01, 732.01, 823.01,1285.01, 1812.01, 1988.01 and 2276.01. It will be interesting to see whether these candidates survive future scrutiny. 

Finally, the right-hand frame of Figure \ref{fig4} shows the distribution of the proposed FPs. Their distribution in the $(P,R)$ plane is fairly uniform and contrasts with both previous plots. Here we have included 158 proposed FPs, which represent less than $7 \%$ of all planetary candidates and $11\%$ of single planet candidates. If the ratio of FPs is much higher, as suggested by studies mentioned earlier in this section, then it is possible that the most or even all of the problematic cases within the sub-Jovian desert may actually be false positives.

\subsection{A Sub-Jovian Pampas}

Even if some of these problematic cases are confirmed as actual planets, the sub-Jovian region with orbital periods below $\sim 2$-$3$ days  still appears significantly underpopulated. In such a case, it would be more appropriate to refer to this region as a {\it sub-Jovian Pampas}, as characterized by a significantly lower planetary occurrence with respect to the surrounding regions. 

Figure \ref{fig5} shows the distribution of confirmed planets and {\it Kepler} candidates according to planetary radii, for three intervals of orbital period (values are specified in the upper right-hand corner of the left plot). For $P > 10$ days, the distribution shows a maximum around $R \sim 2$-$3 R_\oplus$ and sharp decrease in planetary occurrence for both larger and smaller radii. This distribution is not corrected with respect to observational bias, so the real distribution of planets must be significantly different, especially for $R \sim R_\oplus$ (Mayor et al., 2011, Howard et al. 2012). 

The observed distribution for lower orbital periods ($P \in [3,10]$ days) appears bimodal, with a global maximum near $2 R_\oplus$ and a second (local) maximum for Jovian masses. While some Hot Jupiters may have reached his orbital distance through gas disk-driven orbital migration, at least some of them are believed to be the consequence of tidal circularization from high-eccentricity orbits caused by Kozai-capture (e.g. Naoz et al. 2011) or planetary scattering (Nagasawa et al. 2008, Beaug\'e \& Nesvorn\'y 2012). The bi-modality of the planetary distribution for $3 \le P < 10$ days in Figure \ref{fig5} could be because this plot combines discoveries from two different sources (RV surveys and {\it Kepler}) which have different sensitivities and focus on different stellar populations. However, it could also be real, indicating that tidal capture is not as effective for sub-Jovian planets. 

For $P < 3$ days, the bi-modality is even more pronounced, and the sub-Jovian region for these short periods appears severely underpopulated. Again, some (or most) of the Hot Jupiters could have been tidally captured, while most of the Super-Earths could have been driven very close to the star by disc-planet interactions. It is not clear why neither appears to have been effective for sub-Jovian bodies. 

Youdin (2011) presented an analysis of the distribution of {\it Kepler} candidates in the $(P,R)$ plane, fitting different power laws for four sub-samples: planets smaller or larger than $3 R_\oplus$, and orbital periods lower or higher than $P=7$ days. He found significant differences in the size distribution of planets for $P<7$ days and $P>7$ days. This result is closely related to the paucity of sub-Jovian planets discussed here. However, since he did not include data other than {\it Kepler}'s, and did not eliminate possible FPs, none of their distributions exhibited bi-modality.

\section{Distribution of Close-in Planets with Stellar Metallicity}

The paucity of Hot Jupiters (HJ) in detections by {\it Kepler}, with respect to RV surveys, is believed to be due to low metallicity in most of the KIC. Nevertheless, the jury is still out with respect to small masses. Previous spectroscopic analysis of host stars with planets have been limited to those detected with RV surveys (e.g. Santos et al. 2004, Fischer \& Valenti 2005, Johnson et al. 2010, Sousa et al. 2011), and thus mainly to giant planets. These results indicate that Jupiter-size bodies are more likely to be found around metal-rich stars, at least in what concerns the population of planets in close-in orbits. This tendency is not so clear for Neptune-size planets (Sousa et al. 2008, Ghezzi et al. 2010, Sousa et al. 2011), that seem to be found for a wider metallicity range. However, since RV surveys have only been able to detect very few planets in the terrestrial mass range, there has been no clear understanding of the metallicity relation for the occurrence of small planets.

\subsection{Metallicities in the $(P,R)$ Diagram}

This problem was recently undertaken by Buchhave et al. (2012) who analyzed metallicity values for a total of 152 KIC stars. Since the metallicities estimated by {\it Kepler} are photometric and not very precise, Buchhave et al. recalculated some values from very precise spectroscopic measurements. Typical errors for their measurements are of the order of $\sim 0.08$. 

Buchhave et al. (2012) find that stellar metallicities are very diverse for small planets ($R < 4 R_\oplus$), with values between $-0.6$ and $0.5$ with a mean close to ${\rm [m/H]} \sim -0.1$. On the other hand, for larger planets values extend from $-0.2$ and $0.5$ and an average of ${\rm [m/H]} \sim 0.15$. These results indicate that while giant planets appear to require high metallicities, smaller planets can form even around very metal-poor stars. These findings are in agreement with similar results by Udry \& Santos (2007), Sousa et al. (2011) and Adibekyan et al. (2012).

Buchhave et al. (2012) also found that stellar metallicity is not correlated with the orbital distance of planets. However, they focused on distances of a few tenths of AU, and not on the region closer to the star. In fact, from their Figure 2 of the supplementary material, it appears that the region with semimajor axis $a < 0.05$ AU does show a difference with respect to the larger distances. 

\begin{figure}[th]
\centering
\includegraphics[width=0.9\textwidth,clip=true]{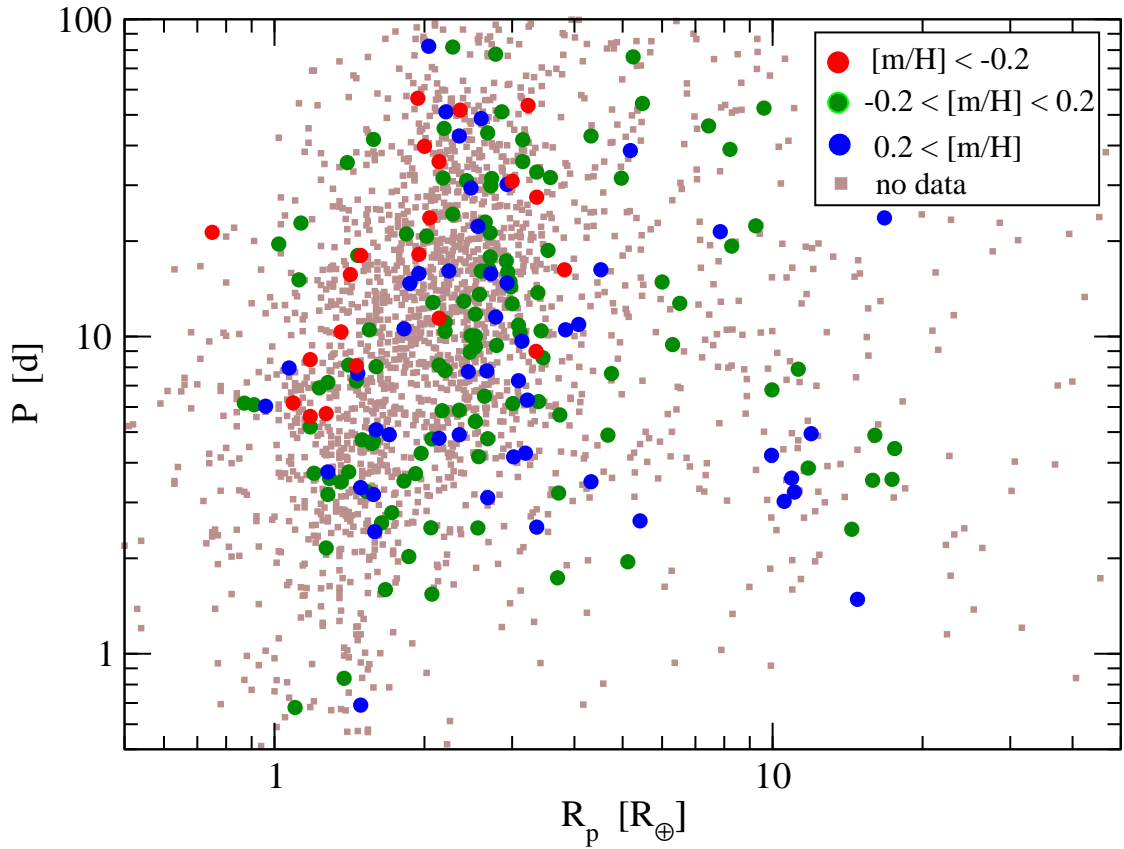}
\caption{Distribution of {\it Kepler} candidates in the $(P,R)$ diagram according to their [m/H] metallicity. Data from Buchhave et al. (2012). Color code is defined in the inset.} 
\label{fig6}
\end{figure}

We used the data kindly provided to us by Lars Buchhave to analyze the metallicity distribution for close-in planets. Results shown in Figure \ref{fig6} are separated into three intervals: metal-poor, Solar metallicity and metal-rich stars. 

The most interesting trend that can be noted in Figure \ref{fig6} is not in the sub-Jovian mass range, but for small planets ($R < 4 R_\oplus$). Small planets belonging to metal-poor stars are located beyond $P > 5$ days, while small planets closer to the star tend to have higher metallicities. 

Another interesting trend from Figure \ref{fig6} is the absence of planets with $R > 4 R_\oplus$ in metal-poor stars. Thus, it appears that in order to form giants or sub-giants, at least a solar metallicity is necessary. 

The trends discussed above could be tied to a more pronounced planetary migration (nebular gas disk or planetesimal driven) in systems with a larger solid content, while small planets formed around metal-poor stars may stay near their formation locations or migrate a lesser amount. Scattering among the small planets could also have played a role. Stars with a higher solid content could tend to form systems of more rocky planets, which would be stable only when their eccentricities are damped by friction (gas or dynamical). Once this stabilization mechanism disappears, close encounters between the planets could lead in some cases to tidal capture and circularization. 

\begin{figure}[t]
\centering
\includegraphics[width=0.65\textwidth,clip=true]{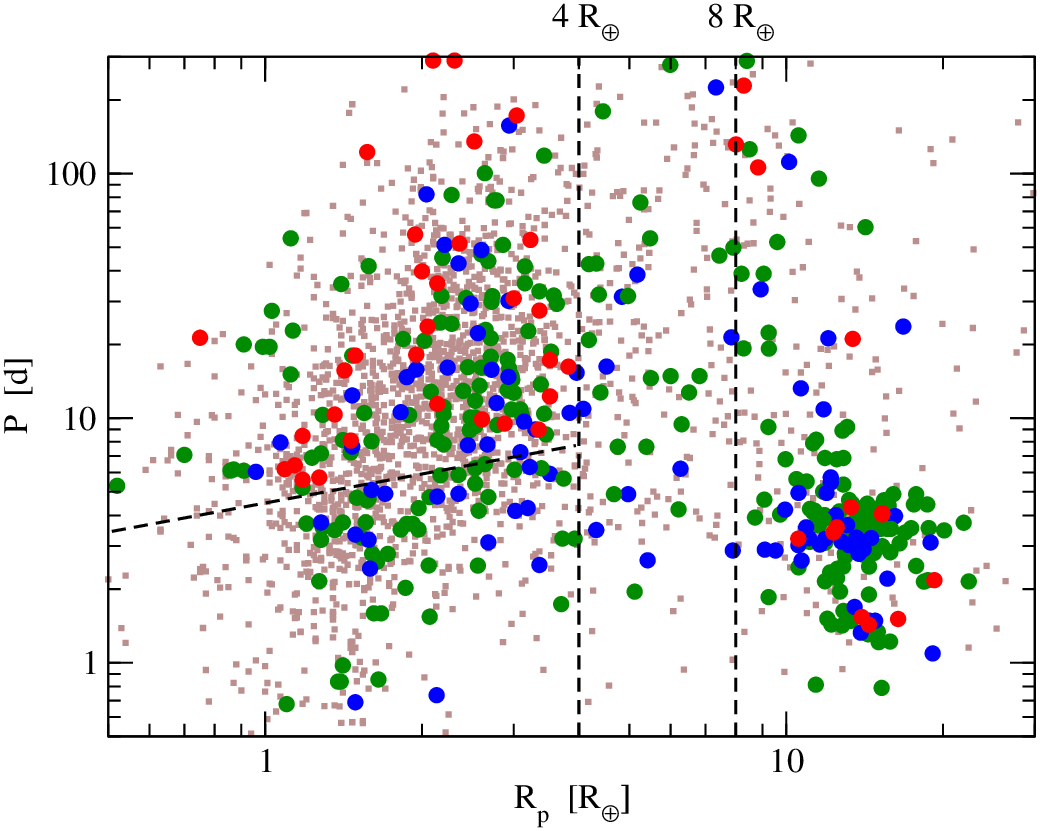}
\caption{Combined sample of stellar metallicities in the $(P,R)$ diagram, including [m/H] data from Buchhave et al. (2012) and spectroscopic [Fe/H] values from RV surveys (Fischer \& Valenti 2005, Sousa et al. 2009, 2011). Color code is same as Figure \ref{fig6}: planets with no available metallicity are shown in small brown squares. Metal-poor stars (${\rm [X/H]} < -0.2$) are shown in red, metal-rich stars (${\rm [X/H]} > 0.2$) in blue, and stars with solar-type metallicities in green. The diagonal dashed line drawn for small planets is indicative of the lower limit found for metal-poor systems.} 
\label{fig7}
\end{figure}

A different data set of metallicities has been obtained for RV detections and/or confirmations. Fischer \& Valenti (2005) and Sousa et al. (2008, 2011) give [Fe/H] spectroscopic values for almost 600 stars with detected planets. Typical errors are of the order of $\sim 0.05$. The advantage of this source is that it would allow us to incorporate most of the currently known Hot Jupiters into the metallicities study and thus have a wider range of planets to analyze. However, we must stress caution, since most of the stars belonging to this data set are located in the solar neighborhood and constitute a different sample form that observed by {\it Kepler}.

The combined metallicities from both samples are shown, in the $(P,R)$ diagram, in Figure \ref{fig7}. A comparison with Figure \ref{fig6} shows very similar trends. The paucity of small ($R < 4 R_\oplus$) planets in metal-poor stars with $P < 5$ days is maintained, even though the number of data points has increased significantly. Actually, the lower limit for metal-poor systems appears to be a diagonal line with orbital periods between $3$ and $6$ days depending on the planetary radius.

For larger masses, we now observe a number of Jovian planets around metal-poor stars, some of them part of the HJ population in the vicinity of the so-called 3-day pile-up. However, there seems to be a curious lack of Sub-Jovian planets (roughly with $4 < R < 8 R_\oplus$) in metal-poor stars for any given orbital period. 

To test the statistical significance of these trends, we once again performed a series of Monte Carlo simulations. For the small planets, we identified in Figure \ref{fig7} the subset of planets with $0.5 R_\oplus \le R \le 4 R_\oplus$ and absolute values of metallicities was larger than $0.1$. We did not consider planets with $|{\rm [X/H]}| < 0.1$ so that the resulting data set had a more or less uniform distribution in metallicities. This gave us a set of $N=99$ planets.

We then ran a series of $10^6$ simulations, in which each body was given a new ${\rm [X/H]}$ value chosen randomly between $-0.5$ and $0.5$ (avoiding absolute values below $0.1$) with an uniform distribution function. We counted what number of these synthetic populations had no fictitious planets with ${\rm [X/H]} < -0.2$ and $P \le 5$ days. The results showed that less than $0.01 \%$ of the cases reproduced the observed trend. 

For intermediate-size planets, we employed the same process. Although the size of the working population was now smaller ($N=21$), we placed no limit on the orbital period. Thus, we now counted what percentage of the test runs had no metallicity value ${\rm [X/H]} < -0.2$. Once again the result were suggestive showing that the observed trend was only reproduced in less than $0.01 \%$ of the cases.

\subsection{Metallicities in the $(P,m)$ Diagram}

Since we have so far worked in the period vs. planetary radius plane, we have only considered detected exoplanets with transit data. This includes both systems with RV and transit, and systems with only transits ({\it Kepler}, CoRoT, etc.). We therefore did not analyze planets for which only RV data is available and, therefore, have undetermined radius.

One way to include these planets in our study is to plot their distribution in the $(P,m\sin{I})$ plane. Not only does this increase the size of our sample, but also allows the use of metallicities determined from RV surveys. The down side is that most of these planets have undetermined orbital inclinations with respect to the line of sight; consequently the masses are minimal values. 

Metallicity data was obtained from Fischer \& Valenti (2005), and Sousa et al. (2008, 2011), and contain values for almost 600 planet hosting stars. Typical errors are of the order of $\sim 0.05$. In particular, Fischer \& Valenti (2005) give estimates for five different elements (Fe, Si, Ti, Na and Ni). The difference between them is of the order of $\sim 0.08$. We chose to use [Fe/H] in order to keep the same indicator as presented for HARPS (Sousa et al. 2008, 2011). 

\begin{figure}[th!]
\centering
\includegraphics[width=0.7\textwidth,clip=true]{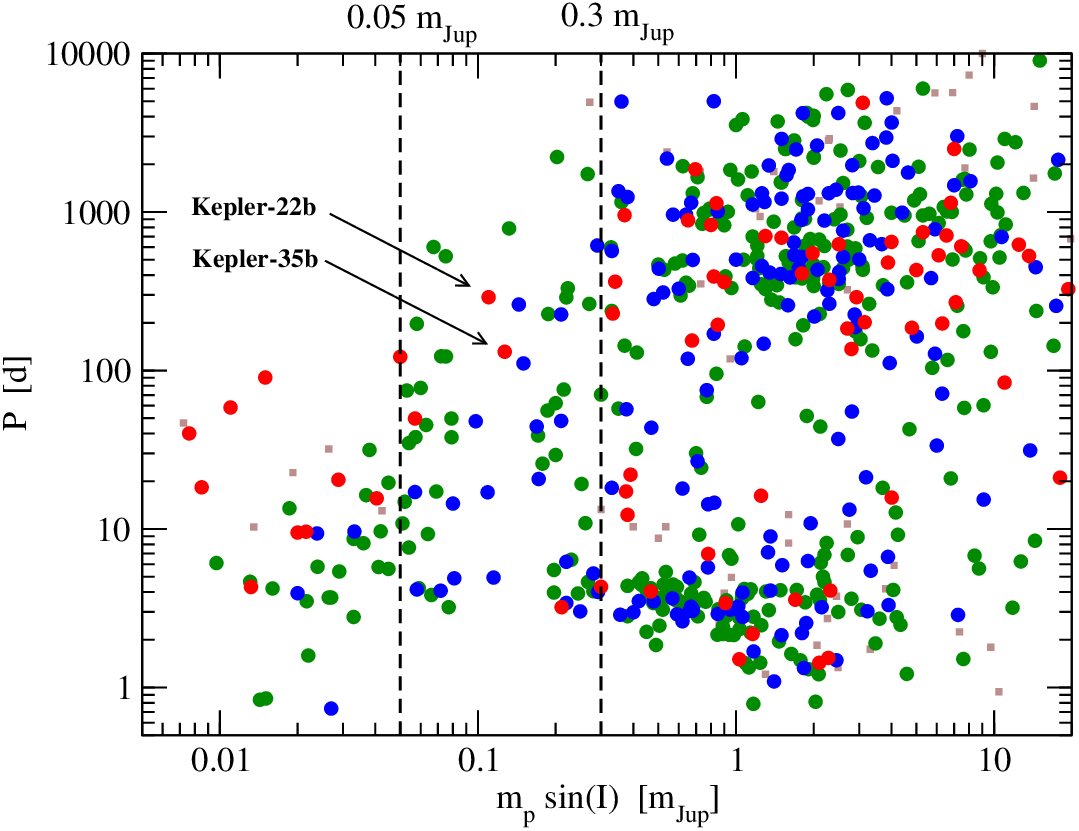}
\caption{Distribution of stellar metallicities in the $(P,m\sin{I})$ plane for confirmed planets with known planetary mass (either true or minimal). Color code is the same as Figures \ref{fig6} and \ref{fig7}: metal-poor stars in red, solar-type in green and metal-rich in blue. The vertical dashed lines indicate rough limits for small planets ($m < m_{\rm Nep} \simeq 0.05 m_{\rm Jup}$) and giant planets ($m > m_{\rm Sat} \simeq 0.8 m_{\rm Jup}$). See text for details.} 
\label{fig8}
\end{figure}

The available data has been summarized in Figure \ref{fig8} and shows very similar trends as detected in Figures \ref{fig6} and \ref{fig7} for the $(P,R)$ plane. For small planets (i.e. $m \sin{I} < 0.05 m_{\rm Jup}$) once again we note that bodies around metal-poor stars are preferably found with larger orbital periods than their metal-rich counterparts. 

For intermediate masses ($m_{\rm Nep} < m \sin{I} < m_{\rm Sat}$), while their distribution in the $(P,R)$ plan showed no planets with ${\rm [x/H]} < -0.2$, Figure \ref{fig8} shows 3 cases, two of which are identified in the plot. Kepler-25b is a planet orbiting a stellar binary in a circumbinary orbit (Welsh et al. 2012) and its formation or evolutionary track could be very different from that of planets around single stars. Kepler-22b is the most distant planet {\it Kepler} has detected so far, with an orbital period of $P = 289$ days. Transit data allow for a fairly precise determination of its radius and correspond to a small planet ($R = 2.3 R_\oplus$, Borucki et al. 2012). Its mass, however, is not well known. Preliminary values, estimated from 16 RV data points, give $m \sim 0.11 m_{\rm Jup}$, which would indicate a sub-Jovian planet. Both values are not easy to reconcile. However, given the small number of RV observations, we believe the location of this planet in the sub-Jovian domain is currently 
questionable. 

The third planet in the sub-Jovian region of the $(P,m\sin{I})$ plane is HAT-P-12b (Hartman et al. 2009), a planet with both RV and transit determinations. HAT-P-12b has a mass of $m = 0.21 m_{\rm Jup}$ and $P \sim 3$ days, placing it barely within the sub-Jovian range (arbitrarily defined), and a radius $R = 11 R_\oplus$, implying the smallest planetary density ($\rho \sim 0.3$ g/cm$^3$) known to date. 

Summarizing, it appears that there are practically no detected Sub-Jovian planets with metallicities below -0.2. This could imply that these bodies are uncommon, or that they are located beyond $P \sim 100$ days, just as in our own Solar System.

\section{Discussion}

We have shown new evidence for a significant paucity of planetary bodies with radii roughly between $3$-$10 R_\oplus$ and orbital periods below $\sim 3$ days. This region is completely void of confirmed planets and {\it Kepler} multi-planet candidates, and was christened by Szabo \& Kiss (2011) as a sub-Jovian desert. However, approximately 16 single-planet {\it Kepler} candidates are located within this region of the $(P,R)$ plane. We find that at least 7 of them are probably false positives. Since we cannot rule out the rest, we prefer to refer to this region as a sub-Jovian Pampas.

The origin of this Pampas is not obvious. It could be related to the effect of atmospheric evaporation (Youdin 2011) which is expected to be especially effective in planets with large gas envelopes and low surface gravity. Very close to the star, atmospheric evaporation would not be effective in planets with high surface gravity (such as Jovian bodies) but could readily strip the volatiles from smaller planets leaving behind the solid cores. In consequence, while most of the Hot Jupiters would not be significantly affected, the observed radius of smaller planets would decrease over time leading to a depletion of this region. 

Although Youdin (2011) only proposes such a mechanism for relatively small planets ($R \sim 3$-$5 R_\oplus$), it may be applicable to a larger interval. Extrapolating from his idea, the depletion of Hot Neptunes would not be complete if sub-Jovian planets originally have very diverse core sizes (relative to their gas envelopes). The change in the planet radius due to atmospheric evaporation would then not be equally effective for all of them. The result would then be a {\it partial} depletion of the region, causing the appearance of the observed sub-Jovian Pampas. However, it is difficult to estimate whether this effect would be effective even up to planetary radii close to Jovian values. 

Another possibility is dynamical in nature. In Beaug\'e \& Nesvorn\'y (2012) we showed that a dynamical tide model (e.g. Ivanov \& Papaloizou 2011) for quasi-parabolic orbits is necessary in order to allow tidal trapping sufficiently far from the central star to avoid stellar engulfment. Dynamical tides, however, are expected to be inefficient for planets with most of its mass in solids (as opposed to gas-rich planets such as Jupiter). Consequently, if the sub-Jovian planets have large cores and light atmospheres, then dynamical tides would not have being able to tidally trap the planets sufficiently far from the star to avoid tidal engulfment, leading to a sparsity of such planets close to the star. 

With respect to the distribution as function of stellar metallicity, the lack of Super-Earths with small orbital period around metal-poor stars may point to a delayed formation of these planets, implying a smaller radial range of planetary migration. The paucity of planets with $R \simeq 4$-$8 R_\oplus$ around metal-poor stars with orbital periods up to $100$ days is also interesting, and could indicate that Neptunes around metal-poor stars did not migrate far and are all located beyond $100$ days. 

A word of caution at this point. We have assumed that the metallicity index [Fe/H] is a proxy for planetary formation. Gonzalez (2009) points out that abundance of other heavy elements (Mg, Si), which together with Fe define the so-called refractory index ``Ref'', could also be important. In their recent survey of chemical abundances for 1111 FGK stars from the HARPS GTO planet search program, Adibekyan et al. (2012) indicate higher [Ref/H] values for Neptune-size planets than [Fe/H] alone. However, the role and relative importance of different refractory materials is not yet firmly established, so it is unclear how using [Ref,H] instead of [Fe,H] could affect our results.

The trends pointed out in this paper are preliminary and we believe they deserve future scrutiny. Future planetary detections and confirmations should be able to validate (or rule out) these trends and allow for a better interpretation of their origin.

\acknowledgements {\bf Acknowledgments:} This work has been supported by a National Science Foundation (NSF) through a Astronomy and Astrophysics Grant (AAG), the Argentina Research Council -CONICET- and the Universidad Nacional de C\'ordoba. We are grateful to Lars Buchhave for kindly providing us with stellar metallicity for many KOI stars. We would also like to thank G. Torres, N.C. Santos and an anonymous referee for critical reading of a previous version of this work and for important suggestions that improved our analysis. Finally, C.B. would also like to express his gratitude to the Southwest Research Institute and Observatoire de la C\^ote d'Azur for invaluable help during the development of this work.

\end{document}